# PROFILS TYPES JOURNALIERS DES CONCENTRATIONS EN PARTICULES DANS LES ENCEINTES FERROVIAIRES SOUTERRAINES PARISIENNES

V. Rakotonirinjanahary[1,4], S. Crumeyrolle[2], M. Bogdan[3], B. Hanoune[1]

[1] Univ. Lille, CNRS, UMR 8522 – PC2A – Physicochimie des Processus de Combustion et de l'Atmosphère, F-59000 Lille, France
[2] Univ. Lille, CNRS, UMR 8518 – LOA – Laboratoire d'Optique Atmosphérique, F-59000 Lille, France
[3] AREP L'hypercube, 16 avenue d'Ivry, 75013 Paris, France
[4] SNCF Holding, 2 Place aux Etoiles, 93210 Saint-Denis, France
*Courriel de l'orateur : miadanavalisoa.rakotonirinjanahary@univ-lille.fr

**TITLE**

TYPICAL DAILY PROFILES OF PM CONCENTRATIONS IN PARISIAN UNDERGROUND RAILWAY STATIONS

**RESUME**

Dans l'objectif d'améliorer la compréhension de la qualité de l'air au sein des enceintes ferroviaires souterraines (EFS), une méthodologie a été développée pour établir un profil de référence des concentrations en particules ($PM_{10}$ et $PM_{2.5}$). Cette approche intègre un processus de nettoyage de données poussé, basé sur l'identification des périodes d'exploitation de l'EFS, des données physiquement incohérentes ou mathématiquement aberrantes. La polyvalence de cette méthodologie permet son application à différentes classes de particules au sein de diverses EFS. Les résultats obtenus à partir des trois EFS étudiées indiquent la possibilité d'obtenir des profils types journaliers fiables même sur des périodes de mesures courtes (jusqu'à une ou deux semaines).

**ABSTRACT**

To enhance the understanding of air quality within underground railway stations (URS), a methodology has been developed to establish a baseline profile of particle concentrations ($PM_{10}$ and $PM_{2.5}$). This approach incorporates an extensive data cleaning process based on the identification of URS operation periods, physically inconsistent, or mathematically aberrant data. The versatility of this methodology allows its application to different particle classes within various URS. The results obtained from the three studied URS indicate the possibility of obtaining reliable daily typical profiles even over short measurement periods (up to one or two weeks).

**MOTS-CLES :** qualité de l'air, enceintes ferroviaires souterraines, concentrations en particules, profil de référence fiable, évolution temporelle de courte et longue durée / **KEYWORDS :** air quality, underground railway stations, PM concentrations, reliable reference profile, short and long-term temporal evolution

## 1. INTRODUCTION

Les trains sont aujourd'hui le moyen de transport privilégié, devenant indispensables pour les déplacements quotidiens en milieu urbain. L'importance de la qualité de l'air à l'intérieur des enceintes ferroviaires souterraines (EFS) a augmenté en raison des espaces confinés et de la forte densité de passagers. Le rapport de l'ANSES (2015) stipule que les concentrations de $PM_{10}$ et $PM_{2.5}$ dans les EFS sont non seulement très supérieures à celles de l'air extérieur mais aussi aux normes de l'OMS (Heinecke, 2021). De plus, Raut et al. (2009) ont trouvé que les concentrations de $PM_{10}$ et $PM_{2.5}$ dans les EFS de Paris étaient environ 5 à 30 fois plus élevées qu'en extérieur, soulignant les préoccupations pour la santé publique. Face à ces préoccupations, la SNCF (Société Nationale des Chemins de Fer français) vise actuellement à installer des systèmes de filtration au sein de ses EFS. Des études comme celles menées par Chang et al. (2021) et Pretot et al. (2022), ont proposé des solutions pour évaluer l'efficacité des systèmes de filtration. L'évaluation de l'efficacité de ces systèmes ne peut se faire que par comparaison avec des périodes sans filtration (situation de référence). La caractérisation de ces situations de référence est rendue difficile par suite de la variabilité temporelle à court et long termes (jour/année), et par les larges fluctuations des mesures. Nous proposons dans ce travail une méthodologie pour déterminer cette situation de référence.

## 2. MATERIELS ET METHODES

### 2.1. Instrumentation et données

Cet article se base sur des données expérimentales provenant de trois EFS parisiennes du réseau SNCF résumées dans le Tableau 1. Pour ces trois EFS, les concentrations de PM ($PM_{10}$ et $PM_{2.5}$) sont mesurées par deux TEOM (Tapered Element Oscillating Microbalance) situés au milieu du quai. Un premier instrument mesure la fraction PM10, l'autre la fraction PM2.5. Les mesures sont prises au pas de temps quart-horaire.

Tableau 1. Caractéristiques des données utilisées

| EFS | EFS 1 | EFS 2 | EFS 3 |
|---|---|---|---|
| Source | Airparif | Airparif | Agence d'Essai Ferroviaire |
| Ligne du RER | C | C | B |
| Type de mesure | Mesure en continue | Campagne de 3 semaines | Mesure en continue |
| Période considérée | 12/04/18 – 31/12/22 | 18/04/17 – 09/05/2017 | 01/01/2021 – 31/12/21 |

La méthodologie a été développée sur l'EFS 1, puis appliquée aux deux autres EFS. La Figure 1 présente l'évolution temporelle des concentrations de $PM_{10}$ dans l'EFS 1 entre avril 2018 et décembre 2022. Cette figure indique également les périodes où des évènements ont perturbé la circulation normale des trains, donc non représentatives du fonctionnement normal et des niveaux de concentrations normaux : les grèves (bandes vertes), les confinements liés à la COVID-19 (bandes orange), les travaux d'été CASTOR (bandes bleues), et les expériences de dépollution des particules (bandes grises). Une tendance relativement constante s'observe tout au long de la période observée, avec des valeurs plus élevées pendant l'été (jusqu'à 500 µg/m$^3$) et des valeurs plus basses pendant l'hiver (jusqu'à 300 µg/m$^3$). Des niveaux de $PM_{10}$ beaucoup plus faibles sont mesurées lors de périodes notables répertoriées ci-dessus, avec par exemple des niveaux maximaux de l'ordre de 100 µg/m$^3$ pendant les confinements COVID, et ont été enregistrés pendant les périodes de grèves (jusqu'à 60 µg/m$^3$) et les confinements liés à la COVID-19 et grève.

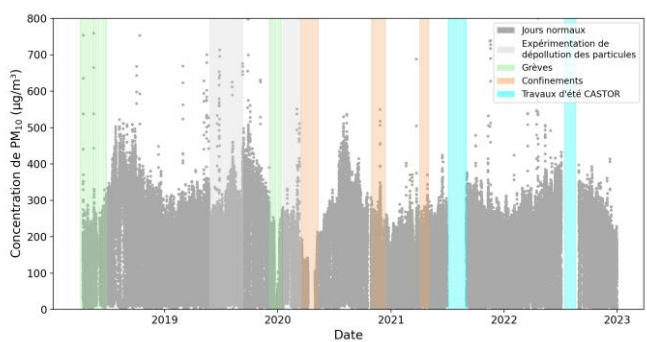

Figure 1. Variation temporelle des concentrations de $PM_{10}$ (12/04/2018 - 31/12/2022) à l'EFS 1

### 2.2. Nettoyage de données

Le nettoyage préliminaire des données peut être décomposé en trois étapes : l'exclusion des périodes notables, des données incohérentes ($PM_{10}$<$PM_{2.5}$), et points aberrants, identifiés en utilisant l'algorithme non supervisé de machine learning, Isolation Forest (Liu et al., 2008). Le Tableau 2 offre un aperçu détaillé du processus de nettoyage des données, mettant en avant les étapes spécifiques utilisées et les proportions correspondantes de points de données retirés de la base de données initiale à chaque étape pour les cas de l'EFS 1. Il est important de noter que la plus grande proportion de points de données retirés est attribuée à l'exploitation de la station, représentant environ 26% de la base de données initiale, tandis que seulement 7% sont liés aux données brutes. Au total, 33,01% des points de données ont été exclus de la base de données initiale.

Tableau 2. Nettoyage des données dans le cas de l'EFS 1

| Etapes de nettoyage des données | Proportion de points de données retirés de la base de données initiale |
|---|---|
| Grèves | 3,98% |
| Confinements liés à la COVID-19 | 7,55% |
| Travaux d'été CASTOR | 5,53% |
| Expérimentation de dépollution des particules | 9,33% |
| Validation des données | 3,97% |
| Détection des anomalies | 2,65% |
| Total | 33,01% |

### 2.3. Méthodologie utilisée pour l'établissement d'un profil type des concentrations en particules

La méthodologie utilisée pour établir le profil type de la concentration en PM est résumée par le diagramme sur la Figure 2. Dans cette méthodologie, les jours ouvrés d'une part et les weekends et jours fériés d'autre part ont été traités séparément vu qu'il y a une exploitation différente de la gare pendant ces deux périodes.

Un premier profil journalier moyen est calculé en utilisant tous les jours restant après le nettoyage, en moyennant point par point au pas de temps quart horaire. Puis chaque jour individuel est comparé à ce profil moyen. Pour cette comparaison, un modèle de régression linéaire forcé à 0 est mis en place dans le but d'identifier les profils individuels (variable dépendante) qui s'écartent du profil moyen journalier (variable indépendante).

En ajustant linéairement chaque profil journalier à ce profil moyen, le coefficient de détermination $R^2$ correspondant nous permet de filtrer les profils individuels qui s'écartent du profil moyen journalier. Dans ce travail, une valeur seuil de $R^2$ égale à 0,75 a été arbitrairement choisie pour éliminer tous les profils individuels éloignés du profil moyen. La pente de régression de cet ajustement, correspond au coefficient d'amplitude journalier.

Enfin, les profils individuels présentant un $R^2$ inférieur à 0,75 sont éliminés, et un nouveau profil moyen est alors recalculé.

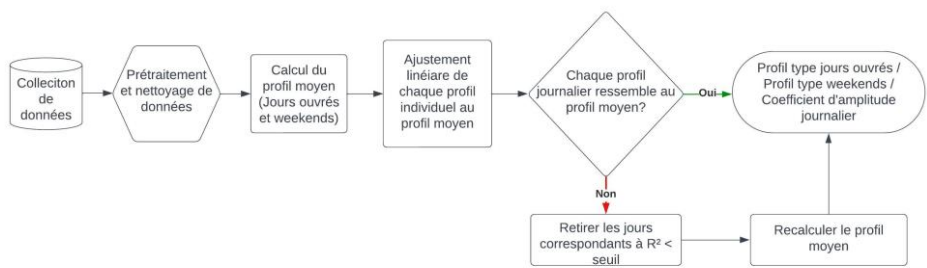

Figure 2. Diagramme pour l'établissement du profil type pour une EFS donnée

Par conséquent, cette méthodologie aboutit à trois résultats exploitables : un profil type pour les jours ouvrés, un profil type pour les week-ends, et un coefficient d'amplitude journalier associé à chaque profil journalier pour chaque jour de la période considérée.

## 3. RESULTATS ET DISCUSSIONS

La Figure 3 présente les profils types journaliers obtenus avec cette méthodologie, pour (a) EFS 1, (b) EFS 2 et (c) EFS 3 tant en jours ouvrés que pendant les weekends. Les barres d'erreurs correspondent à un écart-type. Les résultats révèlent des profils types similaires en semaine et le week-end. Les profils indiqués ont été calculés sur 5 ans pour EFS 1, 3 semaines pour EFS 2 et 1 an pour EFS 3. Néanmoins, nous avons établi, dans les de EFS 1, que même sur une période de 1 semaine, il était possible d'obtenir un profil fiable, s'approchant pour chaque point quart horaire à mieux que 10% du profil établi sur 5 ans. Il est donc pertinent de comparer des obtenus sur des durées différentes.

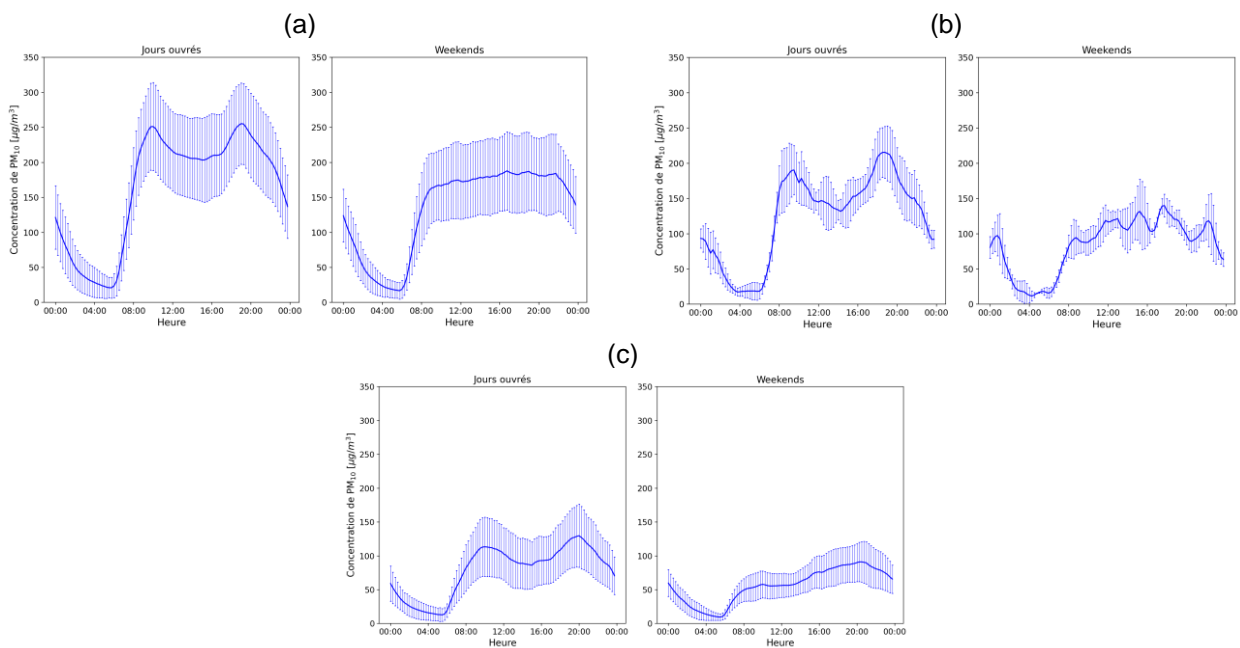

Figure 3. (a) Profils types journaliers aux (a) EFS 1, (b) EFS 2 et (c) EFS 3 pour les jours ouvrés et les weekends

Pour tous ces profils, les concentrations décroissent de la fin du service jusqu'à atteindre environ 10 µg/m$^3$ le matin vers 5 heures, indiquant clairement que la présence de particules est fortement liée à l'exploitation de la gare. Effectivement, l'étude menée par Walther et al. (2017) a montré une forte dépendance entre le mouvement du train et la concentration en particules.

Les résultats révèlent des profils types similaires différents suivant la période (jours ouvrés/weekend) et les gares. En semaine, dans chacune des EFS, deux pics distincts sont observés aux heures de pointe du matin et du soir, avec un pic du soir un peu plus marqué pour les EFS 2 et 3. L'amplitude maximale du profil dépend de la gare, avec un maximum de l'ordre de 250 µg/m$^3$ pour EFS 1(pic du soir), 200 µg/m3 pour EFS 2, et 100 µg/m3 pour EFS 3. Le weekend, les profils obtenus dans les trois EFS sont différents, dans leur allure et leur niveau maximal. Ces comportements montrent qu'en plus de la fréquence des trains, d'autres facteurs sont à prendre en compte, tel que les caractéristiques spécifiques de la gare (forme, profondeur de la gare, système de ventilation…), son environnement extérieur, ainsi que la météo (Chang et al., 2021).

Il est à noter que ces profils peuvent servir de référence de base, non seulement pour évaluer les expériences d'amélioration de la qualité de l'air, mais aussi comme référence pour des analyses ultérieures.

## 4. CONCLUSIONS

Cette étude présente une nouvelle méthodologie pour établir un profil de référence au sein d'une EFS. Dans une première étape, un nettoyage standardisé des données est effectué (exclusion des périodes d'exploitation spécifique, des données incohérentes, et des outliers). Dans une deuxième étape, une méthode itérative de moyennage est mise en place pour aboutir au profil de référence sur toute la période considérée. Chaque profil journalier est alors transformé en un coefficient d'amplitude plus facilement exploitable. Des résultats très satisfaisants sont obtenus même pour des périodes d'analyse courtes, jusqu'à une ou deux semaines.

Les EFS étudiées montrent des différences significatives dans leurs profils types, qui doivent maintenant être analysés en fonction des divers paramètres identifiés dans ce travail et la littérature.

Cette méthodologie sera appliquée à l'étude de l'efficacité de systèmes de filtration de particules installés dans les EFS.

## 5. REFERENCES ET REMERCIEMENTS